\documentclass[aps,prd,preprint,showpacs,preprintnumbers,superscriptaddress,nofootinb]{revtex4}
\usepackage{epsfig,epstopdf}
\usepackage{amsfonts,amsmath,amssymb,bm,dsfont,eucal}
\usepackage[english]{babel}
\usepackage[latin1]{inputenc}
\usepackage[T1]{fontenc}
\usepackage{ae,aecompl}
\usepackage{graphicx,graphics,subfigure,color,psfrag}
\usepackage{hyperref}
\newcommand{\be}{\begin{equation}}
\newcommand{\ee}{\end{equation}}
\newcommand{\bea}{\begin{eqnarray}}
\newcommand{\eea}{\end{eqnarray}}

\newcommand{\as}{\alpha_{\mathrm{s}}}

\begin{document}
\addtolength{\jot}{10pt}
\addtolength{\jot}{10pt}

\title{\bf On the QCD result  for  the hyperfine splitting $M_{\Upsilon(1S)}-M_{\eta_b}$\\
and the value of  $\as$}

\author{Pietro Colangelo}
\affiliation{Istituto Nazionale di Fisica Nucleare, Sezione di Bari, Italy}
\author{Pietro Santorelli}
\affiliation{Dipartimento di Scienze Fisiche, Universit\`a di Napoli ``Federico II'',  Napoli, Italy}
\affiliation{Istituto Nazionale di Fisica Nucleare, Sezione di Napoli, Italy}
\author{Egidio Scrimieri}
\affiliation{Dipartimento di Fisica, Universit\`a di Bari,   Italy}
\affiliation{Istituto Nazionale di Fisica Nucleare, Sezione di Bari, Italy}

\vspace*{3cm}
\begin{abstract}
\noindent The  measurement of the $\eta_b$ mass,  together with   a QCD
result for the hyperfine splitting  $E_{HFS}=M_{\Upsilon(1S)}-M_{\eta_b}$,  allows us to  determine
 the strong coupling  constant $\as$ at a low energy scale. The result
\be
\as(M_{\Upsilon(1S)})=0.197\pm0.002  \big|_{\Delta E_{HFS}^{exp}} \pm 0.002  \,
\big |_{scheme} \pm 0.002  \,
\big |_{\delta <G^2>}  \pm 0.006  \, \big |_{ \delta m_b}\, \pm 0.005  \, \big |_{\mathrm{ho}} \,\,\,\, , \nonumber
\ee
\be
\displaystyle
\as(M_{Z^0})=0.124\pm0.001\, \big|_{\Delta E_{HFS}^{exp}} \pm 0.001  \,  \big |_{scheme}
\pm 0.001  \,
\big |_{\delta <G^2>}  \pm 0.003  \, \big |_{ \delta m_b}\, \pm 0.002  \, \big |_{\mathrm{ho}} \,\,\,  \nonumber
\ee
\noindent
is compatible with the current world average of $\as$ reported by the Particle Data Group,
and  shows that the experimental lowest-lying $\bar b b$ hyperfine splitting can be reproduced in terms
of a perturbative and nonperturbative QCD contribution.
\end{abstract}
\vspace*{2cm}
\preprint{BARI-TH-621-09}
\preprint{DSF-2009-12}
\maketitle

\newpage
The observation of the $\eta_b$ by the BaBar \cite{Babar:2008vj,BaBar:2009pz}
and CLEO Collaborations \cite{cleo}
comes after  three decades of  searches of the lightest pseudoscalar $b \bar b$ meson.
The spin-singlet state $\eta_b$ has been detected  in the radiative $\Upsilon(3S)\to \eta_b \gamma$
and $\Upsilon(2S)\to \eta_b \gamma$ decay modes,  studying the spectrum of the final photon.
The measured mass  reported by  the BaBar Collaboration  is
\begin{equation}
M_{\eta_b} = 9388.9^{+3.1}_{-2.3}\, (\mathrm{stat}) \pm 2.7\, (\mathrm{syst})\, \mathrm{MeV}
\end{equation}
\noindent
from  $\Upsilon(3S)\to \eta_b \gamma$  \cite{Babar:2008vj},  and
\begin{equation}
M_{\eta_b} = 9394.2^{+4.6}_{-4.8}\, (\mathrm{stat}) \pm 2.0\, (\mathrm{syst}) \, \mathrm{MeV}
\end{equation}
\noindent
from  $\Upsilon(2S)\to \eta_b \gamma$   \cite{BaBar:2009pz}.

The signal  of $\eta_b$ in the $\Upsilon(3S)\to \eta_b \gamma$ radiative decay
has been confirmed by the CLEO Collaboration,
which quotes \cite{cleo}
\begin{equation}
M_{\eta_b} = 9391.8 \pm 6.6\, (\mathrm{stat}) \pm 2.0\, (\mathrm{syst}) \, \mathrm{MeV} \,\,\, .
\end{equation}
\noindent

The three mass measurements, combining in quadrature the statistic
and systematic uncertainties,  produce  the average value  \cite{PDG}
\begin{equation}\label{eq:mass-exp}
M_{\eta_b}= 9390.9\pm 2.8\, \mathrm{MeV} \,\,\,
\end{equation}
which is a remarkable result, since it provides us with a measurement of the
hyperfine splitting (HFS) of the lowest-lying $\bar b b$ doublet,
\begin{equation}\label{eq:HFS-exp }
E_{HFS}^{exp} = M_{\Upsilon(1S)}-M_{\eta_b}= 69.3\pm 2.8 \, \mathrm{MeV}\,\,\, ,
\end{equation}
where we used the Particle Data Group value  $M_{\Upsilon(1S)}=9460.30 \pm 0.26$ MeV
for the mass of $\Upsilon(1S)$  \cite{PDG}.

The experimental result  (\ref{eq:HFS-exp })  can be compared to the predictions of
 quark models \cite{Giannuzzi},
 lattice QCD \cite{Gray:2005ur}, and QCD sum rules \cite{QCDSR}.
 Moreover, it is particularly interesting since it can be compared to the
 result of evaluations based on perturbative QCD with the inclusion of the
 leading nonperturbative contribution,
 an expression involving fundamental QCD parameters such as the strong coupling
 constant $\as$ at a low energy scale. The
 obtained value of $\as$ can be compared to other determinations, and considered
 when the average is carried out.
 In this way, one can also investigate if there is room, in the experimental result,
 for contributions not related to QCD, such as that from the mixing effect
 envisaged in \cite{lozano} under the assumption of the existence of a light $CP$-odd pseudoscalar Higgs.
 The determination of $\as$ from the result  (\ref{eq:HFS-exp }) is the purpose of the present study.

As recognized since the early studies of  quantum chromodynamics
applied to mesons comprising heavy quarks
\cite{old},
a quantitative description in QCD of the   $Q \overline{Q}$ bound state  is possible if
the average distance between the quark pair  is smaller than the typical QCD length scale
$r \simeq 1/\Lambda_{QCD}$; the description is given in terms of  a perturbative
QCD expression and nonperturbative corrections.

The perturbative contribution to HFS comes from diagrams with external heavy
quark-antiquark lines and the exchange of gluons and light-quark loops.
At the leading order in the $\alpha_s$ expansion, such diagrams produce,
in the static limit,  the Coulombic  $Q \overline{Q}$ potential
\begin{equation}
V^{(0)}_{Q \bar Q}=- C_F\frac{\alpha_s}{r}
\end{equation}
with $C_F=(N_c^2-1)/(2N_c)$ the eigenvalue of the quadratic Casimir operator
of  the fundamental representation  of the group $SU(N_c)$,    $N_c$  being the number of colors.
The expression of the potential at  one- \cite{pot-one} and two-loop \cite{pot-two} orders
has been recently enlarged to three loops \cite{pot-three};
the meson masses are obtained as energy levels of a Schr\"odinger equation.

At the leading order in $\alpha_s$,  the perturbative contribution to the
hyperfine splitting is proportional  to the beauty quark mass and to the fourth power of $\as(\mu)$:
 \begin{equation}\label{eq:LO }
E_{HFS}^{LO} = \frac{C_F^4 \as^4(\mu) m_b}{3} \,\,\,\ .
\end{equation}
In this expression,  the dependence  of $\as$ on the renormalization scale $\mu$ requires
a proper choice of this parameter in order to apply the formula to the  physical case.
A milder $\mu$  dependence can be achieved including higher order corrections.
The  $O(\as)$ corrections to the $E_{HFS}^{LO}$ leading order term have been computed in
Refs. \cite{titard,penin}, and the result includes  a logarithmically enhanced  $\as \log\as$  term.
Such kinds  of terms can be resummed to all orders through a renormalization group
analysis carried out in the framework of the  (potential) nonrelativistic QCD effective theory,
and indeed in Ref. \cite{Kniehl:2003ap} a Next-to-Leading Log (NLL) expression of the hyperfine splitting has been derived which
includes a resummation of terms of the form $\as^n \log^{n-1}\as$, with
$\as$ renormalized in the $\overline {\mathrm {MS}}$ scheme.
A discussion can be found in \cite{Penin:2009wf}.

In the following we use the formula for $E_{HFS}^{NLL}$ in \cite{Kniehl:2003ap}
(corrected in version 2 of the preprint in the arXiv), together with
 the expression of $\as$ to four loops \cite{Chetyrkin:1997sg}:
\begin{eqnarray}\label{eq:alphas }
&&\as^{(4)}(\mu)=\frac{1}{\beta_0\,L}\,
\left\{1-\frac{\beta_1}{\beta_0^2}\frac{\ln L}{L}+
\frac{1}{\beta_0^2 L^2}\left[\frac{\beta_1^2}{\beta_0^2}
\left(\ln^2L-\ln L-1\right)+\frac{\beta_2}{\beta_0}\right]+\right.\nonumber\\
&&\left.\frac{1}{\beta_0^3L^3}\left[\frac{\beta_1^3}{\beta_0^3}\left(
-\ln^3L+\frac{5}{2}\ln^2L+2\ln L -\frac{1}{2}\right)-3\frac{\beta_1\beta_2}
{\beta_0^2}\ln L+\frac{\beta_3}{2\beta_0}\right]\right\}\,,
\end{eqnarray}
where $L = \ln\left(\mu^2/\Lambda^2\right)$ and  $\beta_i$  are given by
\cite{vanRitbergen:1997va}
\begin{eqnarray}\label{eq:betai }
&&\beta_{0}={1 \over 4\pi} \left[11-\frac{2}{3}n_{f}\right]\nonumber\\
&&\beta_{1}={1 \over (4\pi)^2} \left[102-\frac{38}{3}n_{f}\right]\nonumber\\
&&\beta_{2}={1 \over (4\pi)^3} \left[\frac{2857}{2}
   -\frac{5033}{18}n_{f}+\frac{325}{54}n_{f}^2\right]\nonumber\\
&&\beta_{3}={1 \over (4\pi)^4} \left[\left(\frac{149753}{6}+
   3564\zeta_{3}\right)-\left(\frac{1078361}{162}+
  \frac{6508}{27}\zeta_{3}\right)n_{f}\right . \nonumber\\
&&\hspace{2.2truecm}+\left . \left(\frac{50065}{162}+\frac{6472}{81}\zeta_{3}\right)n_{f}^2
  +\frac{1093}{729}n_f^3\right] \,\,\,\  ;
\end{eqnarray}
$n_f$ is the number of active flavors ($n_f=4$ in the case of the $\bar b b$ system)
and  $\zeta_3=\zeta(3)$.
The renormalization group improved expression of  $E_{HFS}^{NLL}$ involves the beauty
quark mass $m_b$ as an overall factor. It also involves the strong coupling
$\as$ evaluated at a low energy renormalization scale $\mu$ and
at a matching scale $m_b$,  as well as on the QCD parameter $\Lambda$.
The dependence on the two different scales  $\mu$ and $m_b$ allows us, through an error analysis, to bound the value of $\Lambda$.

For a heavy $Q \overline{Q}$ pair the nonperturbative contribution to the hyperfine splitting is related to the dynamics
of the colored quarks in the gluon background. If the  size of the quarkonium system
is smaller than the fluctuations of the background
gluon field, this background  field can be considered homogeneous and constant, and
parametrized by  gluon condensates.
The lowest dimensional  nonperturbative contribution to the hyperfine splitting of
the lightest  $\bar b  b$ doublet has been evaluated in \cite{Voloshin:1978hc,Leutwyler:1980tn}
and involves the  dimension-four gluon condensate,
$\displaystyle <G^2>= \langle 0| \frac{\alpha_s}{\pi} G^a_{\mu \nu}ÊG^{\mu\nu a}|0\rangle$:
\begin{equation}
\label{nonpert}
E^{NP}_{HFS} = \frac{m_b}{3} \left( C_F^4 \alpha_s \tilde \alpha_s^3 \right)
\frac{18.3  \, \pi^2 <G^2>}{m_b^4 (C_F \tilde \as)^6}  \,\,\, .
\end{equation}
In this  expression the coupling $\widetilde{\alpha}_s$ is defined as
\begin{eqnarray}
\widetilde{\alpha}_s(\mu) & = & \alpha_s(\mu)
\left\{
1+\left(a_1+\gamma_E\frac{\tilde \beta_0}{2}\right)\frac{\alpha_s(\mu)}{\pi}+\right. \nonumber\\
&&\left. \left[\gamma_E\left(a_1\tilde \beta_0+\frac{\tilde \beta_1}{8}\right)+\left(\frac{\pi^2}{12}+
\gamma_E^2\right)\frac{\tilde \beta_0^2}{4}+a_2\right]
\frac{\alpha_s^2(\mu)}{\pi^2}
\right\}\,,\,\,
\end{eqnarray}
with $\tilde \beta_i=(4 \pi)^{i+1} \beta_i$,
since  the effective Coulombic potential $\displaystyle V_{eff}^0=-  C_F \frac{ \tilde \as}{r}$
has been considered with the inclusion of  two-loop corrections  \cite{titard,Yndurain:1999ui}.
The parameters $a_1$ and $a_2$ are given by
\begin{eqnarray}
a_1 & = & \frac{31\, C_A-20\,T_F\, n_f}{36} \nonumber\\
a_2 & = & \frac{1}{16}\left\{\left[\frac{4343}{162}+4\pi^2-\frac{\pi^4}{4}+\frac{22}{3}\zeta(3)\right]C_A^2\right.\nonumber\\
&&\left. -\left[\frac{1798}{81}+\frac{56}{3}\zeta(3)\right]C_A\,T_F\,n_f\right.\nonumber\\
&& \left.-\left[\frac{55}{3}-16\zeta(3)\right]C_F\,T_F\,n_f+\frac{400}{81}T_F^2\,n_f^2\right\} \,\,\, ,
\end{eqnarray}
with $\gamma_E$ the Euler constant, $C_A=N_c$ the eigenvalue of the quadratic Casimir
operator of  the adjoint  representation  of  $SU(N_c)$, and $T_F=\frac{1}{2}$.
This contribution must be added to the perturbative one, and represents the first term,
in  the vacuum condensate expansion,  of a series
involving condensates of higher dimension  and higher powers of the inverse heavy
quark mass  \cite{Pineda:1996uk}.

In the  theoretical expression  of  $E_{HFS}$  the coupling constant  $\as$ appears both in the
perturbative and nonperturbative terms, and  the formula
\begin{equation}
E_{HFS}= E^{NLL}_{HFS} + E^{NP}_{HFS}
\end{equation}
involves the factor $m_b$, the renormalization and the matching scales, together with the QCD parameter
$\Lambda^{(n_f=4)}$,  the  QCD scale in our problem with four active flavors.
We fix  the gluon condensate to  the commonly accepted value
$\displaystyle \langle 0| \frac{\alpha_s}{\pi} G^a_{\mu \nu}ÊG^{\mu\nu a}|0\rangle=(0.012\pm 0.004)$ GeV$^4$
\cite{Colangelo:2000dp},  and we include  the uncertainty on the value of $m_b$ in the denominator of (\ref{nonpert})
 in the uncertainty of the condensate.
In the same uncertainty, which is of about $33 \%$,  we can also include the effect
of a possible difference in the scale of $\alpha_s$ in the non perturbative contribution with respect to the
perturbative one.
 At odds with other analyses,   we keep the nonperturbative contribution  using its face value,
 instead of, e.g.,  fixing it from the charmonium hyperfine splitting and
then rescaling to the bottomonium case \cite{Kniehl:2003ap}.

Proceeding in the numerical analysis,  we divide $E_{HFS}^{exp}$ and $E_{HFS}$ by
$M_\Upsilon/2$, obtaining
$\tilde E_{HFS}^{exp}$ and $\tilde E_{HFS}$. In the QCD expression, this amounts to dividing by the mass
$m_b$ defined in the $1S$ bottom quark mass scheme.
The change of the scheme induces a higher order $\alpha_s$ correction in the theoretical formula,
which is beyond the chosen level of accuracy;   in $\tilde E_{HFS}$ the remaining $m_b$
dependence is only logarithmic, and encodes the dependence of the matching scale.
Hence, the procedure is to evolve $\alpha_s$ from $M_{Z^0}$ to $m_b$ with  four-loop accuracy,
and  below $m_b$ to use $\alpha_s$
according to the logarithmic precision of the expression for  $E_{HFS}$.  To check  the uncertainty
in this procedure,  we have also used full four-loop accuracy for $\alpha_s$
at all scales; the  difference in the numerical result  is included in the final error budget.

\begin{figure}[t]
\begin{center}
\includegraphics[width=0.45\textwidth]{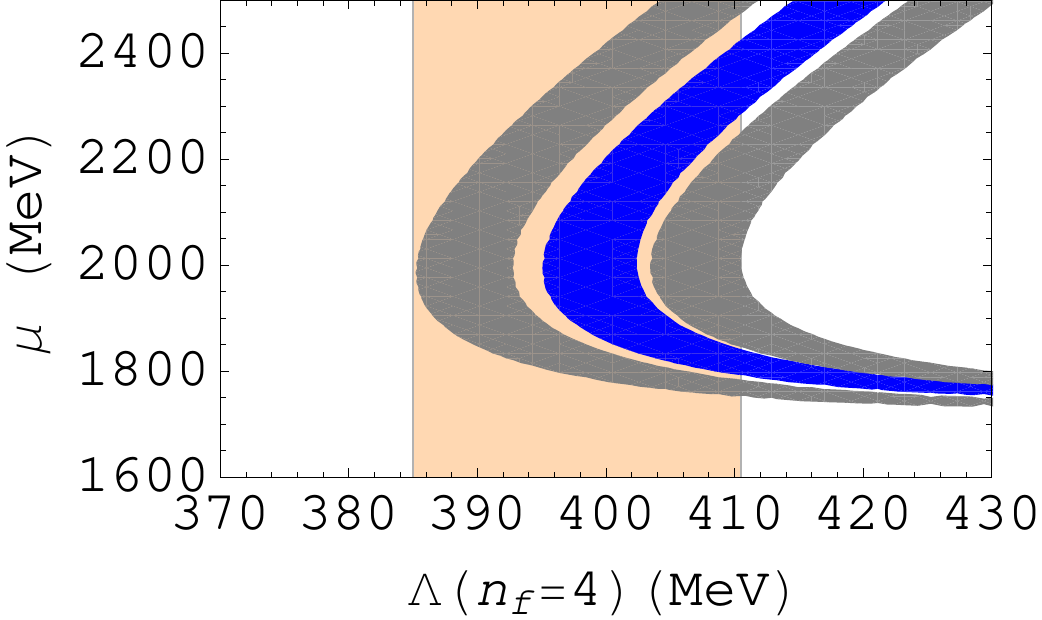}
\end{center}
\caption{\baselineskip=15pt Correlation between the renormalization scale $\mu$ and $\Lambda^{(n_f=4)}$  from the $\bar b  b$
hyperfine splitting at a fixed matching scale.   The central (blue) curve refers to
the central value of $\tilde E^{exp}_{HFS}$, the left and  right
(gray) curves  to $\tilde E^{exp}_{HFS} - \Delta  \tilde E^{exp}_{HFS}$ and
$\tilde E^{exp}_{HFS} + \Delta  \tilde E^{exp}_{HFS}$ , respectively, and the
matching scales are $m_b=4724$,  $4746$ and $4691$  MeV, respectively.
To minimize the dependence of  $\Lambda^{(n_f=4)}$ on $\mu$, a vertical
band is found through the condition $\displaystyle \frac{\partial \Lambda^{(n_f=4)}}{\partial \mu}=0$.
}
\label{correlation-b}
\end{figure}

We define the function
\begin{equation}\label{dat}
\xi(\Lambda,\mu,m_b)=\frac{(\tilde E_{HFS}^{exp}-\tilde E_{HFS}^{NLL}-\tilde E_{HFS}^{NP})^2}{(\Delta \tilde E_{HFS}^{exp})^2}
\end{equation}
where $\Lambda=\Lambda^{(n_f=4)}$,  and search its minima in the space of parameters $\mu$, $\Lambda$ and $m_b$.
To fix the range of $\Lambda$ allowed by the comparison between the experimental datum and
the theoretical expression we proceed in the following way.
The minima of $\xi(\Lambda,\mu,m_b)$  determine an implicit relation between its variables.
Once we have obtained a set of three values  $(\Lambda^*,\mu^*, m_b^*)$
corresponding to the minimum of $\xi$, we fix $m_b$ to the value $m_b^*$ and study
$\xi(\Lambda,\mu, m_b^*)$ as a function of $\Lambda$ and $\mu$
(the central band depicted in Fig.\ref{correlation-b}). The  procedure is repeated using in the
expression of  $\xi$, as the experimental datum, the values
$\tilde E^{exp}_{HFS} - \Delta  \tilde E^{exp}_{HFS}$ and
$\tilde E^{exp}_{HFS} + \Delta  \tilde E^{exp}_{HFS}$ respectively, obtaining
the left and right bands in Fig. \ref{correlation-b},  hence bounding the parameters $\Lambda$ and
$\mu$ to a region of the parameter plane depicted in Fig.\ref{correlation-b}.

\begin{figure}[t]
\begin{center}
\includegraphics[width=0.45\textwidth]{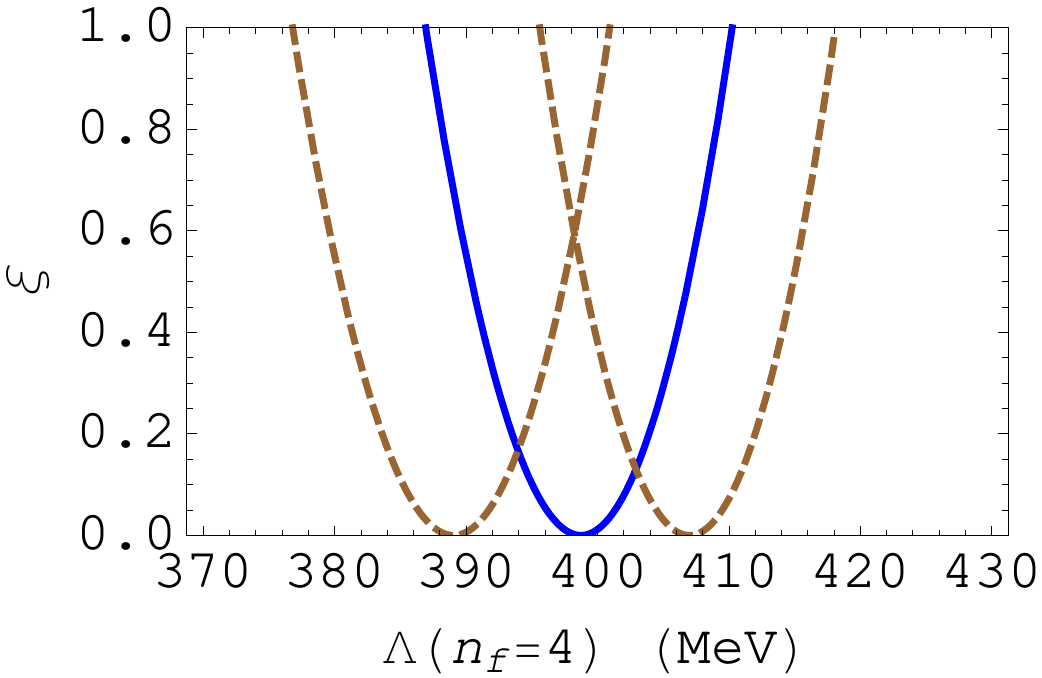}
\end{center}
\caption{\baselineskip=15pt The function $\xi(\Lambda,\mu,m_b)$ at $\mu=2$ GeV  for the three cases
$\tilde E^{exp}_{HFS}$ (continuous blue curve) and
$\tilde E^{exp}_{HFS} \pm \Delta  \tilde E^{exp}_{HFS}$
(dashed brown curves). The values of $m_b$ are fixed as
in Fig.\ref{correlation-b}.}
\label{lambda4}
\end{figure}

Along the curves of minima, there are ranges of $\Lambda^{(n_f=4)}$ where the dependence on the
renormalization scale $\mu$ is minimized.  We bound  $\Lambda^{(n_f=4)}$
in these ranges,  imposing that the condition
$\displaystyle \frac{\partial \Lambda^{(n_f=4)}}{\partial \mu}=0$ is satisfied:
the corresponding band is depicted in Fig.\ref{correlation-b}.
In this way we bound   $\mu$  within  $1800  \, {\rm  MeV} \leq \mu   \leq 2200    \, {\rm  MeV}$ (slightly larger values than the scale
$\mu\simeq1500$ MeV chosen in  \cite{Kniehl:2003ap} to compute the central value of $E_{HFS}^{NLL}$),   and we obtain  for $\Lambda^{(n_f=4)}$,
\begin{equation}
\Lambda^{(n_f=4)} = 398^{+12}_{-13}  \,\,\,\,  {\rm MeV}  \,\,\, .
\end{equation}
At $\mu=2000$ MeV the values of $\Lambda^{(n_f=4)}$ where the function $\xi$ vanishes are depicted
in Fig.\ref{lambda4}. In this region of the parameter space, the perturbative contribution to
the hyperfine splitting amounts to $E_{HFS}^{NLL}=65.84$ MeV,  while the nonperturbative
contribution is $E_{HFS}^{NP}=3.58$ MeV; therefore, for the
lowest-lying beauty doublet  the splitting is mainly of perturbative origin.
 It is interesting to consider the case where the non perturbative term is forced to be zero.
 In this condition, the experimental value of $E_{HFS}^{exp}$ is reproduced through  a larger value
of the QCD parameter:  $\displaystyle \Lambda^{(n_f=4)} = 414^{+10}_{-13}$  MeV, so that
an uncertainty of $16$ MeV can be attributed to
 $\Lambda^{(n_f=4)}$ from the  $D=4$ gluon condensate. Moreover, using four-loop accuray in $\alpha_s$ at all scales or
 the procedure of using $\alpha_s$ with different accuracies described above induces an error of
 $\pm 20$ MeV on  $\Lambda^{(n_f=4)}$, quoted as $\Delta \Lambda^{(n_f=4)}|_{scheme}$
   in the final result.

In the $m_b-\Lambda$ parameter plane, a simple correlation is found between the parameters,
which can be easily understood due to the logarithmic dependence
of $E_{HFS}^{NLL}$
on $m_b/\Lambda$.  The uncertainty on $\Lambda$ is linked to
the variation of the matching scale, which cannot be sensibly larger than
$M_\Upsilon/2$.  A variation of $250$ MeV of this scale induces a shift of
$\Lambda^{(n_f=4)}$  of about $40$ MeV, an uncertainty  dominating the error of $\Lambda^{(n_f=4)}$.
The central value of the result for $\Lambda^{(n_f=4)}$ corresponds to $m_b=M_\Upsilon/2$.

The last source of uncertainty comes from the neglect of  (uncalculated) higher order
contributions to $E_{HFS}$.  The size of these contributions has been estimated
considering the difference $E_{HFS}^{NLL}-E_{HFS}^{LL}$, with the conclusion that
it is  about $20.5\%$ of the central value of $E_{HFS}^{NLL}$
  \cite{Kniehl:2003ap}. This effect  produces  an uncertainty, quoted as $\mathrm{ho}$ (higher orders),
  of  $\pm33$ MeV to $\Lambda^{(n_f=4)}$. To be conservative, we include this uncertainty in the final error,
even though some higher order effects (for example in the accuracy of $\as$) have been considered separately.

Following all the  steps in the outlined procedure,  we obtain a result for $\Lambda^{(n_f=4)}$
from the  experimental $\bar b  b$ hyperfine splitting:
\begin{equation}
\Lambda^{(n_f=4)} = 398^{+12}_{-13} \,  \Big|_{\Delta E_{HFS}^{exp}} \pm20 \Big|_{scheme} \pm 16 \,
\Big |_{\delta <G^2>} \pm 40 \, \Big |_{ \delta m_b}  \,  \pm33 \, \Big |_{\mathrm{ho}} \,\,\,\,  {\rm MeV}  \,\,\, .
\label{lambda-result}
\end{equation}
%

With the value  of $\Lambda^{(n_f=4)}$ in (\ref{lambda-result}) it is possible to evolve $\alpha_s$
to  the $\Upsilon(1S)$ and   to the $Z^0$  mass scale, implementing the proper matching condition
at $\mu  = M_f$ to include the fifth flavor,
the beauty \cite{Chetyrkin:1997sg,Prosperi:2006hx}:
\begin{equation}
\as^{(n_f-1)}(M_f)=\as^{(n_f)}(M_f)\left[1+
k_2\left(\frac{ \as^{(n_f)}(M_f) }{\pi}\right)^2 +
k_3\left(\frac{\as^{(n_f)}(M_f)}{\pi}\right)^3\right]
\end{equation}
with
$\displaystyle k_2=\frac{11}{72}$ and
$\displaystyle k_3=\frac{564731}{124416}-\frac{82043}{27648}\zeta_3-\frac{2633}{31104}(n_f-1)$.
We find
\begin{equation}
\as(M_{\Upsilon(1S)})=0.197\pm0.002  \big|_{\Delta E_{HFS}^{exp}} \pm 0.002  \,
\big |_{scheme}
\pm 0.002  \,
\big |_{\delta <G^2>}  \pm 0.006  \, \big |_{ \delta m_b} \pm 0.005  \, \big |_{\mathrm{ho}}  \label{as-meas1}
\ee
and
\be
\as(M_{Z^0})=0.124\pm0.001\, \big|_{\Delta E_{HFS}^{exp}} \pm 0.001  \,
\big |_{scheme}  \pm 0.001  \,
\big |_{\delta <G^2>}  \pm 0.003  \, \big |_{ \delta m_b}  \pm 0.002  \, \big |_{\mathrm{ho}} \,\,\,\, .
\label{as-meas}
\ee
Equations (\ref{as-meas1}) and (\ref{as-meas})  show  the quality of the  determination of the strong coupling constant
from the $\bar b b$ hyperfine splitting:  for comparison,   the  determination of $\as$
 from the ratio
$R_\gamma=\Gamma(\Upsilon \to \gamma gg)/\Gamma(\Upsilon \to ggg)$
of radiative/hadronic decay widths of $\Upsilon(1S)$  corresponds to
$\displaystyle \as(M_{\Upsilon})=0.184^{+0.015}_{-0.014}$ and
$\displaystyle \as(M_{Z^0})=0.119^{+0.006}_{-0.005}$  \cite{brambilla}.

\begin{figure}[t]
\begin{center}
\includegraphics[width=0.45\textwidth]{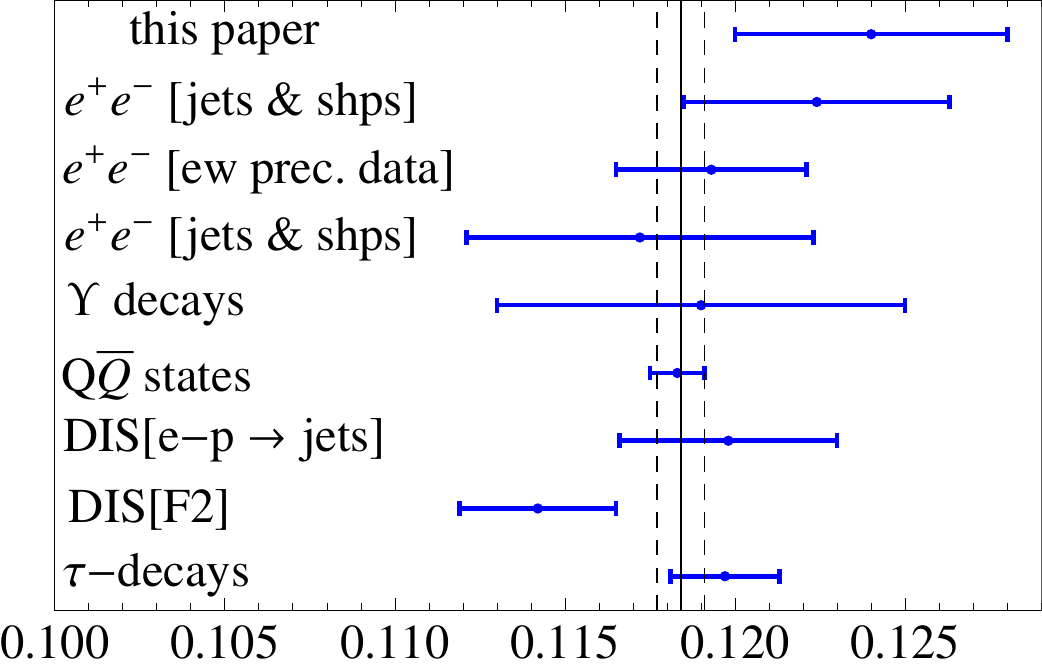}
\end{center}
\caption{\baselineskip=15pt
Measurements of $\alpha_s(M_{Z^0})$ used
in \cite{Bethke:2009jm} to compute the 2009 world average,  together with the determination of
$\alpha_s(M_{Z^0})$ obtained in this paper.
The continuous vertical line corresponds to the world average value in \cite{Bethke:2009jm},
and the dashed lines take the error into account: $ \as(M_{Z^0})=0.1184\pm0.0007$. This
result is dominated by the  HPQCD determination (indicated as $Q \overline Q$ states)
\cite{Davies:2008sw}.}
\label{alphas}
\end{figure}

The result in Eq.(\ref{as-meas}) can be compared to the  world average of
$\as$. The 2009 average computed in \cite{Bethke:2009jm} is obtained considering,
together with the result from  the ratio of radiative/hadronic $\Upsilon$ decay widths,
the determinations of $\as$ from  $\tau$-lepton  decays, deep inelastic scattering
processes (in particular, from nonsinglet structure functions and jet production rates),
$e^+ e^-$ processes (event shapes and jet production rates), and from electroweak precision fits.
Moreover, a determination of  the HPQCD Collaboration,
based on the analysis of the $Q \overline{Q}$ system on the lattice, is included:
$ \as(M_{Z^0})=0.1183\pm0.0008$  \cite{Davies:2008sw}.  As one can see by looking at
Fig. \ref{alphas}, this last value dominates the
present average, $\as(M_{Z^0})=0.1184\pm0.0007$ \cite{Bethke:2009jm}, and
compared to this average, the value quoted in (\ref{as-meas}) is less than $2 \sigma$ higher.
The  result (\ref{as-meas}) from  the $\bar b  b$ hyperfine splitting,
$\alpha_s(M_{Z^0}) = 0.124 \pm 0.004$ (with the error obtained by combining in quadrature the
various uncertainties), together with the 2009 world average of $\as$ obtained in
\cite{Bethke:2009jm}, slightly increases the value:
\begin{equation}
\as(M_{Z^0})=0.1186 \pm 0.0007 \,\,\,\, .
\label{aver}
\end{equation}

Our conclusion is that there is the possibility  to accommodate
the experimental datum on the hyperfine splitting of the lowest-lying
$\bar b b$ doublet with the QCD, perturbative and nonperturbative,
description of it.  The resulting value of $\as$ is compatible
with the world average within less than 2 standard deviations.
The inclusion of the value of $\as$  determined in this paper
slightly increases the average.

\end{document}